\documentclass{article}
\usepackage{spconf,amsmath,graphicx,hyperref}
\usepackage{booktabs}
\usepackage{tabularx}
\usepackage{multirow}
\usepackage{caption}
\usepackage{amsmath} 
\usepackage{amssymb}
\usepackage{adjustbox} 
\usepackage{multirow}
\usepackage{tabularx}
\usepackage{url}
\usepackage{multirow}
\usepackage{booktabs}

\usepackage{array}
\usepackage{verbatim}

\title{DiaMoE-TTS: A Unified IPA-Based Dialect TTS Framework with Mixture-of-Experts and Parameter-Efficient Zero-Shot Adaptation}
\name{%
  \begin{tabular}{c}
    Ziqi Chen$^{1,*}$\\
    Chaofan Ding$^{2}$
  \end{tabular}
  \qquad
  \begin{tabular}{c}
    Gongyu Chen$^{2,*}$\\
    Zihao Chen$^{2,\dagger}$
  \end{tabular}
  \qquad
  \begin{tabular}{c}
    Yihua Wang$^{2}$\\
    Wei-Qiang Zhang$^{1,\dagger}$
  \end{tabular} 
  \thanks{$^{\ast}$ These authors contributed equally to this work. $^{\dagger}$ Corresponding author.}
}

\address{$^{1}$ Department of Electronic Engineering, Tsinghua University \\
         $^{2}$ AI Lab, Giant Network}

\begin{document}
%
\maketitle
\begin{abstract}
Dialect speech embodies rich cultural and linguistic diversity, yet building text-to-speech (TTS) systems for dialects remains challenging due to scarce data, inconsistent orthographies, and complex phonetic variation. To address these issues, we present DiaMoE-TTS, a unified IPA-based framework that standardizes phonetic representations and resolves grapheme-to-phoneme ambiguities. Built upon the F5-TTS architecture, the system introduces a dialect-aware Mixture-of-Experts (MoE) to model phonological differences and employs parameter-efficient adaptation with Low-Rank Adaptors (LoRA) and Conditioning Adapters for rapid transfer to new dialects. Unlike approaches dependent on large-scale or proprietary resources, DiaMoE-TTS enables scalable, open-data-driven synthesis. Experiments demonstrate natural and expressive speech generation, achieving zero-shot performance on unseen dialects and specialized domains such as Peking Opera with only a few hours of data. Code and resources are available at: \url{https://github.com/GiantAILab/DiaMoE-TTS}
\end{abstract}
\section{Introduction}
\label{sec:intro}



Zero-shot text-to-speech (TTS) models have advanced rapidly in recent years\cite{valle,mehta2024matcha,sparktts2503.01710,casanova2022yourtts,e2tts}, demonstrating remarkable capabilities in generating natural and expressive speech. Dialect TTS, a vital extension of speech synthesis, is culturally and linguistically significant across China’s diverse dialects and many regional languages in Europe, yet progress is hindered by data scarcity, inconsistent orthographies, and front-end modeling challenges. Public resources for dialects are largely automatic speech recognition (ASR)-oriented rather than TTS-ready; to our knowledge, there is no unified \emph{open-data} pipeline that enables end-to-end zero-shot dialect TTS.

To address this gap, we leverage linguistic expertise to construct a novel multilingual speech dataset for Chinese dialects. Built upon existing open-source ASR corpora, our dataset introduces a unified International Phonetic Alphabet (IPA)-based pronunciation lexicon\cite{International_Phonetic_Association_1999}, providing aligned text, IPA phonemes, and speech pairs. This resource enables consistent modeling across diverse dialects and serves as a foundation for open-data-driven, zero-shot multilingual and multi-dialect TTS research.

Based on the above, we propose a unified framework for low-resource dialect TTS that leverages both cross-lingual pre-training and phonetically grounded modeling. We standardize the pronunciation modeling across dialects using the IPA, constructing a harmonized text-IPA-speech alignment dataset covering multiple Chinese dialects. 






Building upon the F5-TTS architecture, we introduce three key contributions:  

\noindent \textbullet\ A unified IPA-based front-end that replaces conventional pinyin or character inputs with explicit phonetic sequences, ensuring precise pronunciation modeling for Chinese dialects.  

\noindent \textbullet\ A Mixture-of-Experts (MoE) module in the text embedding layer to capture dialect-aware phonological variations.  

\noindent \textbullet\ A strategy for rapid adaptation to extremely low-resource dialects while retaining knowledge from high-resource ones.  

Our method has reached a performance almost comparable to closed-source or industrial-scale systems. We will also open-source the dataset construction methodology, a multidialect dataset from public data, and full training/inference scripts to support reproducible research.

\section{Related Work}
\subsection{IPA: A Standardized Phonetic Representation}
\begin{figure}[!t] 
  \centering
  \includegraphics[width=0.5\textwidth]{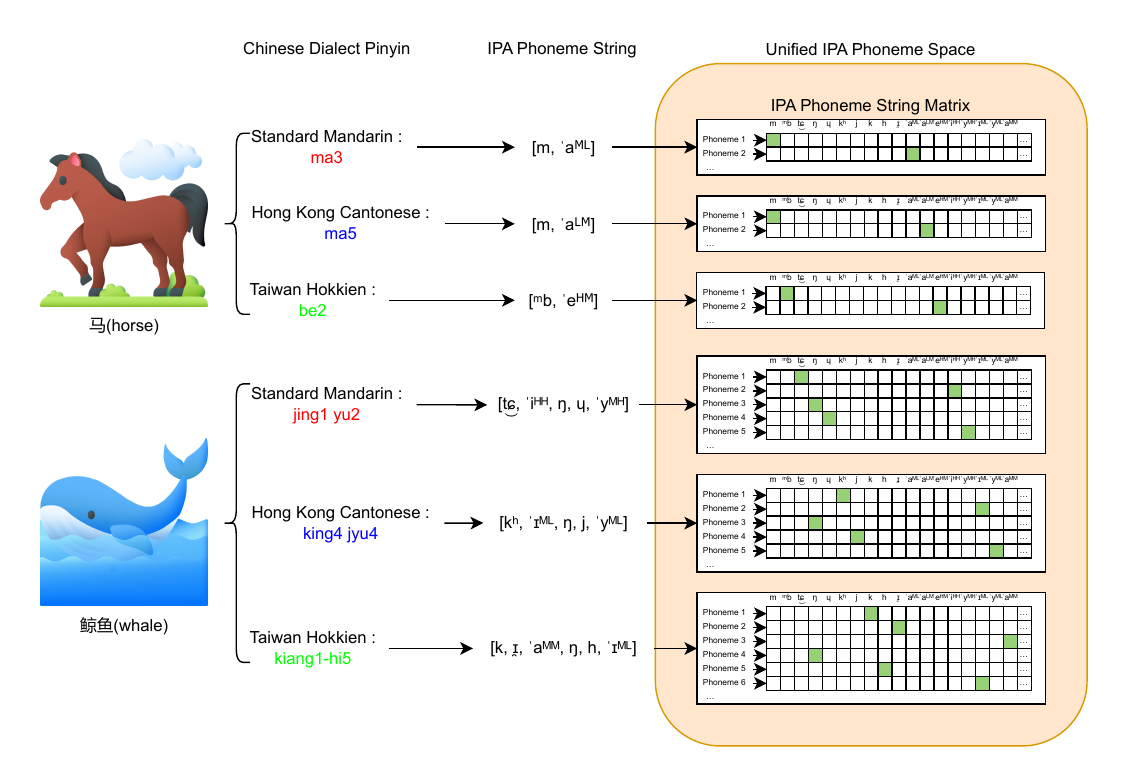} 
  \caption{Unified dialect frontend based on IPA. The same character may vary in phonetic and prosodic forms across dialects, but IPA symbols consistently capture these variations, unlike pinyin.}
  \label{fig:ipa}
\end{figure}

IPA is a standardized phonetic system, enabling precise and unambiguous representation of human speech. In early TTS systems, IPA plays a crucial role\cite{Taylor_2009}, offering accurate phonetic guidance and reducing mispronunciations. Its consistency also makes it essential for building unified multilingual TTS models\cite{ipa_hemati2022usingipabasedtacotrondata,ipa_liu2023tacotron2}. As shown in the figure \ref{fig:ipa}, we have built a unified dialect front-end system based on IPA. Different types of dialect share a set of identifiers, providing a prerequisite for the expansion of low-resource dialect TTS.
\subsection{F5-TTS: A Lightweight and Efficient High-Quality TTS Model}

F5-TTS~\cite{f5tts} is a fully non-autoregressive TTS model built upon \textbf{Optimal Transport Conditional Flow Matching} (OT-CFM)~\cite{lipman2024flowmatchingguidecode} and a \textbf{Diffusion Transformer} (DiT) backbone~\cite{Peebles2022DiT}. It synthesizes speech in the reference speaker’s voice given a reference audio, its transcript, and target text. The model encodes text using ConvNeXt V2 blocks~\cite{woo2023convnextv2}, conditions on masked Mel-spectrograms and flow-matching latents, and employs DiT layers to progressively denoise latent spectrograms, effectively capturing long-range prosody and speech structure.


\section{Approach}
\label{sec:approach}


\begin{figure}[!t] 
  \centering
  \includegraphics[width=0.48\textwidth]{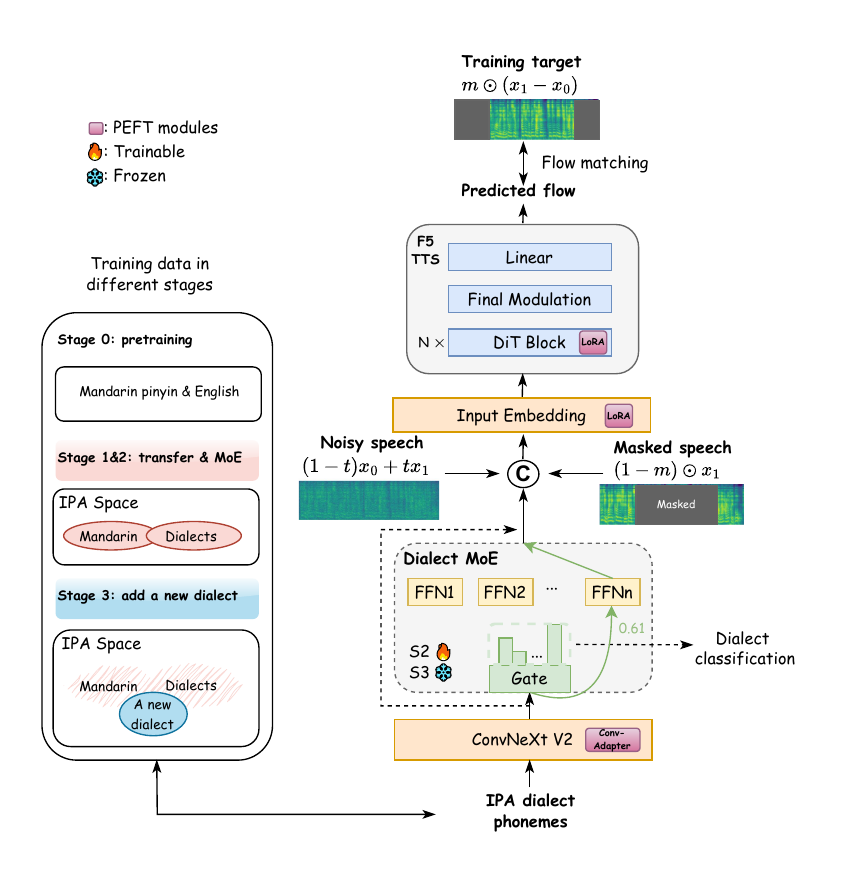} 
  \caption{The overall architecture of the proposed dialect TTS system. In stages 1 and 2, the IPA data of Mandarin and multiple dialects were jointly used for training, and in stage 2, MoE was additionally added. In stage 3, only the new dialect was used to train PEFT modules\cite{loraref1hu2022lora,loraref2kim2024voicetailor}.}
  \label{fig:overview}
\end{figure}

\subsection{Overview}

We propose a unified framework for dialect TTS, built upon a multi-stage training strategy that progressively extends the model's capability from Mandarin to Chinese dialectal speech and finally enables adaptation to low-resource novel dialects. The proposed framework is trained in four stages.

\noindent\textbf{Stage 0: Initialization with F5-TTS checkpoint} \\
The training starts from a pre-trained F5-TTS checkpoint, trained on large-scale Mandarin and English data. This provides a strong prior for high-quality speech generation and serves as the foundation for downstream adaptation.

\noindent\textbf{Stage 1 \& 2: Joint Training on Multidialect IPA Data} \\
In both Stage 1 and Stage 2, the model is trained on a unified dataset of IPA-aligned text-speech pairs covering Standard Mandarin and multiple Chinese dialects. It is designed to align diverse input representations (e.g., pinyin, characters) with a unified IPA phoneme space. In stage 2, we add an extra dialect-style MoE module\cite{moe_xue2025moettsenhancingoutofdomaintext} after Text Embedding to enable the model to automatically learn and capture the relevant information of dialect styles, thereby weakening the phenomenon of style averaging.

\noindent\textbf{Stage 3: Adaptation to Low-Resource Dialects} \\
Building upon the multidialect-capable model, Stage 3 performs rapid adaptation to new, low-resource dialects. We adopt a parameter-efficient fine-tuning (PEFT) approach\cite{pefttts_kwon25_interspeech} in which only LoRA and Conditioning Adapters are trained, while the backbone model remains frozen. Furthermore, data augmentation methods are also introduced to expand the small training set. This enables speech synthesis with only a few hours of data, preserving multilingual knowledge and mitigating catastrophic forgetting.

This staged training paradigm effectively balances model capacity, data efficiency, and generalization, enabling rapid expansion to new dialects in zero-shot or few-shot speech synthesis.
\subsection{Dialect-style MoE in Text Embedding}
While our method employs a single, shared text encoder that assumes relatively stable grapheme-to-phoneme mappings. However, the joint training of multiple dialects can lead to the homogenization of styles, and without additional guidance, the expression of dialectal styles will be weakened. To address this, we incorporate a residual MoE architecture into the text embedding module:

\begin{equation}
h = E_{\text{text}}(p_{\text{IPA}}),\quad h' = \operatorname{MoE}(h) + h
\end{equation}

\noindent where \(p_{\text{IPA}}\) indicates input IPA frontend sequence, \(h\) and \(h'\) are the original and the dialect-aware feature, respectively.

Our MoE module comprises multiple expert networks, each specialized in capturing the phonological characteristics of specific dialects. A learnable gating mechanism dynamically routes input sequences to the most relevant experts based on contextual and linguistic cues. As a guide, we introduce an auxiliary dialect classification loss to encourage the MoE gate. The gate logits are computed from a mean-pooled representation of the input sequence:
\begin{equation}
s = \frac{1}{T} \sum_{t=1}^{T} h_t \in \mathbb{R}^D, \quad
g = \mathrm{Gate}(s) \in \mathbb{R}^K
\end{equation}
where $ h_t \in \mathbb{R}^D $ is the hidden state at time step $ t $, $ T $ is the sequence length, $ K $ is the number of experts (i.e., dialects), and $ \mathrm{Gate}(\cdot) $ is a linear projection followed by optional activation.

Given the gate logits $ \mathbf{g}_i \in \mathbb{R}^K $ for sample $ i $ and the ground-truth dialect label $ y_i \in \{1,\dots,K\} $, the dialect classification loss is computed as the mean-reduced cross-entropy over the batch:
\begin{equation}
\mathcal{L}_{\text{dialect}} = -\frac{1}{N} \sum_{i=1}^{N} \log  \frac{\exp(\mathbf{g}_{i,y_i})}{\sum_{j=1}^{K} \exp(\mathbf{g}_{i,j})} 
\label{eq:cross_entropy}
\end{equation}
where $ N $ is the batch size.
Finally, the total loss becomes:
\begin{equation}
\mathcal{L}_{\text{total}} = \mathcal{L}_{\text{OT-CFM}} + \lambda \cdot \mathcal{L}_{\text{dialect}}
\end{equation}
with $\lambda = 0.1$ during stage 2 and $\lambda = 0$ in other stages.

This enables learning of dialect-aware text representation, allowing the model to adaptively generate prosodic patterns that align with the target dialect. In addition, the sparse activation of experts ensures scalability and computational efficiency, making it feasible to support a wide range of dialects within a unified framework.

\subsection{Low-resource Dialect Adaptation Method}
Due to the scarcity of high-quality speech data for many dialects—often limited to less than ten hours—we further investigate the generalization of our framework to extremely low-resource dialects. Inspired by \cite{pefttts_kwon25_interspeech}, We incorporate a Conditioning Adapter into the text embedding module and apply LoRA to the query and value projections within the attention layers of the input embedding and DiT blocks. The MoE structure is kept frozen during fine-tuning to preserve its learned routing behavior and avoid overfitting under limited data.

We also use acoustic data augmentation techniques to expand the limited training data of the new dialect without altering its dialectal style. Specifically, for each original utterance, we apply pitch and time-scale modifications with factors of 0.85, 0.9, 0.95, 1.05, 1.1, and 1.15. By combining with this simple and effective augmentation method, the zero-shot speech synthesis capability of this model can be extended to a new dialect without affecting the original dialect synthesis ability.

\begin{table}[t]
\centering
\caption{Comparison of Training Data Scale Across TTS Systems}
\label{tab:data_scale}
\small
\renewcommand{\arraystretch}{0.8}
\begin{tabular}{l c}
\toprule
\textbf{System} & \textbf{Data Scale}/hours \\
\midrule
Edge TTS\cite{edge_tts}         & - \\
CosyVoice2\cite{du2024cosyvoice}        & $\sim$150k \\
Qwen-TTS\cite{qwen_tts_2025}         & $\sim$3,000k \\
\textbf{Ours} (stage 1\&2)    & $\sim$\textbf{0.7k} Mandarin+\textbf{0.4k} dialects \\
\bottomrule
\end{tabular}
\end{table}

\begin{table*}[t]  
\centering
\caption{Main Results on WER, MOS and UTMOSv2 Scores Across Dialects}
\label{tab:main_results}
\footnotesize
\renewcommand{\arraystretch}{0.8}
\setlength{\tabcolsep}{5pt}
\begin{tabular}{l l c c c c c c c c c c c}
\toprule
\textbf{Metric} & \textbf{Model} & \textbf{YUE} & \textbf{SH} & \textbf{CD} & \textbf{XA} & \textbf{ZZ} & \textbf{TJ} & \textbf{NAN} & \textbf{SJZ} & \textbf{NJ} & \textbf{Jingbai} & \textbf{Yunbai} \\
\midrule

\multirow{4}{*}{\textbf{WER} /\%} 
    & Ours                  & 76.59 & 76.53 & 37.12 & 33.00 & 29.59 & 20.18 & 92.41 & 36.83 & 31.64 & 39.95 & 68.94 \\
    \cmidrule(lr){2-13}
    & Edge TTS \cite{edge_tts}       & 4.15 & --   & 0.70 & 0.34 & 0.35 & --   & --   & --   & --   & --   & --   \\
    & CosyVoice2 \cite{du2024cosyvoice} & 59.81 & 49.01 & 10.65 & --   & 8.15 & 6.23 & --   & --   & --   & --   & --   \\
    & Qwen TTS \cite{qwen_tts_2025}     & --   & 48.68 & 10.95 & --   & --   & --   & --   & --   & --   & --   & --   \\
    & \textit{Average}                & 31.98 & 48.85 & 7.43 & 0.34 & 4.25  & 6.23 & -- & -- & --   & --   & --   \\
\midrule

\multirow{4}{*}{\textbf{MOS}} 
    & Ours                  & 3.03 & 1.80 & 2.22 & 3.15 & 2.97 & 1.66 & 2.16 & 2.67 & 1.75 & 3.04 & 2.84 \\
    \cmidrule(lr){2-13}
    & Edge TTS \cite{edge_tts}       & 3.70 & --   & 3.50 & 4.01 & 3.68 & --   & --   & --   & --   & --   & --   \\
    & CosyVoice2 \cite{du2024cosyvoice} & 2.76 & 2.22 & 4.06 & --   & 4.07 & 2.64 & --   & --   & --   & --   & --   \\
    & Qwen TTS \cite{qwen_tts_2025}     & --   & 4.43 & 4.58 & --   & --   & --   & --   & --   & --   & --   & --   \\
    & \textit{Average}                & 3.07 & 2.96 & 4.05 & 4.01 & 3.94  & 2.64 & -- & -- & --   & --   & --   \\
\midrule

\multirow{5}{*}{\textbf{UTMOSv2}} 
    & Ours                  & 2.65 & 2.75 & 3.33 & 2.87 & 2.86 & 3.10 & 3.10 & 2.56 & 3.28 & 1.75 & 1.94 \\
    \cmidrule(lr){2-13}
    & Edge TTS \cite{edge_tts}       & 2.56 & --   & 3.19 & 3.18 & 2.93 & --   & --   & --   & --   & --   & --   \\
    & CosyVoice2 \cite{du2024cosyvoice} & 2.65 & 2.99 & 2.83 & --   & 3.08 & 3.12 & --   & --   & --   & --   & --   \\
    & Qwen TTS \cite{qwen_tts_2025}     & --   & 2.77 & 3.31 & --   & --   & --   & --   & --   & --   & --   & --   \\
    & \textit{Average}                & 2.61 & 2.88 & 3.11 & 3.18 & 3.01 & 3.12 & --   & --   & --   & --   & --   \\
\midrule

\textbf{Diff.} 
& Ours vs Comp. Avg. (WER) & 44.61 & 27.52 & 29.69 & 32.66 & 25.34 & 13.95 & -- & -- & -- & -- & -- \\
    & Ours vs Comp. Avg. (MOS) & -0.05 & -1.15 & -1.83 & -0.86 & -0.96 & -0.98 & -- & -- & -- & -- & -- \\
    & Ours vs Comp. Avg. (UTMOSv2)    & 0.04  & -0.13 & 0.22  & -0.31 & -0.15 & -0.02 & -- & -- & -- & -- & -- \\

\bottomrule
\end{tabular}
\end{table*}

\begin{table}[t]
\centering
\caption{Ablation Study on MOS and WER}
\label{tab:ablation}
\footnotesize
\renewcommand{\arraystretch}{0.8}  
\setlength{\tabcolsep}{4pt}
\begin{tabular}{l c c c c c c}
\toprule
\textbf{Model} & \textbf{MoE} & \textbf{Input Rep.} & \multicolumn{4}{c}{\textbf{Dialects}} \\
\cmidrule(lr){4-7}
& & & \textbf{CD} & \textbf{XA} & \textbf{ZZ} & \textbf{SJZ} \\
\midrule
\textbf{MOS} \\
w/o MoE 
    & $\times$ 
    & IPA 
    & \textbf{2.46} & 2.33 & 2.61 & 2.39 \\
w/o IPA
    & \checkmark 
    & pinyin 
    & 1.23 & 1.04 & 1.16 & 1.19 \\
Ours 
    & \checkmark 
    & IPA 
    & 2.22 & \textbf{3.15} & \textbf{2.98} & \textbf{2.67} \\
\midrule
\textbf{WER}/\% \\
w/o MoE 
    & $\times$ 
    & IPA 
    & \textbf{45.08} & 41.09 & 41.54 & 49.38 \\
w/o IPA
    & \checkmark 
    & pinyin 
    & 93.29   & 93.00   & 90.65   & 90.49   \\
Ours 
    & \checkmark 
    & IPA 
    & 49.01   & \textbf{33.00}   & \textbf{29.59}   & \textbf{36.83}   \\
\bottomrule
\end{tabular}
\end{table}
\section{Experiments}
\label{sec:exp}

\subsection{Data}
We utilize the Common Voice Cantonese dataset\cite{commonvoice:2020}, the Emilia Mandarin dataset\cite{emilia}, dialectal data from the KeSpeech corpus\cite{2021KeSpeech} and a open-source Sourthern Min dataset\cite{nandata}. In addition, we incorporate commercially acquired speech data in the Shanghai and Tianjin dialects to broaden the coverage of dialect. For low-resource fine-tuning, we employ approximately 3 hours of professionally recorded Peking Opera recitation, covering both Jingbai and Yunbai styles, along with an equal amount of Nanjing dialect data for comparison. 

\subsection{Implementation Details}
We use the AdamW optimizer for the model in all experiments. In stages 1 and 2, a peak learning rate of 7.5e-5 is employed, with a warm-up of 2k steps and a linearly decayed learning rate setting over 200K total training steps. The batch size is set to 28k frames per GPU. In Stage 3, only the LoRA (rank=16, $\alpha$=1) and Conditioning Adapters are updated, while the main model is frozen. We trained the model for 100K steps with a learning rate of $1 \times 10^{-5}$.




\subsection{Metrics}
We adopt a combined objective and subjective evaluation framework. For objective evaluation, we report word error rate (WER) using FireRedASR\cite{xu2025fireredasr} as the evaluation ASR model; and UTMOSv2\cite{2024utmosv2}, a learned metric for predicting speech naturalness. For subjective evaluation, we conduct human listening tests using mean opinion score (MOS). 
\subsection{Evaluation}
In this study, we evaluate our model on multiple Chinese dialects. The abbreviations used in the tables correspond to the following dialects: YUE (Cantonese), SH (Shanghai), CD (Chengdu), XA (Xi’an), ZZ (Zhengzhou), TJ (Tianjin), NAN (Southern Min), SJZ (Shijiazhuang), NJ (Nanjing), Jingbai and Yunbai (two pronunciation types of Peking Opera).
\subsubsection{Comparison with Existing Methods}
This evaluation covers both objective metrics and subjective listening tests, comparing our system against several state-of-the-art commercial TTS models including Edge TTS\cite{edge_tts}, CosyVoice2\cite{du2024cosyvoice}, and Qwen-TTS\cite{qwen_tts_2025}.

Tables~\ref{tab:main_results} shows our method’s performance across all dialects, including low-resource varieties like Peking Opera (Jingbai and Yunbai) and Nanjing. Our approach effectively learns from limited data via continual learning, whereas direct fine-tuning leads to significant forgetting of prior knowledge.

Table \ref{tab:main_results} also displays the WER, MOS and UTMOSv2. While our system falls slightly behind commercial baselines in certain metrics, this observation warrants further clarification. Notably, the primary objective of our method is not to surpass highly optimized proprietary systems, but rather to demonstrate a feasible framework for training with minimal resources. In addition, the apparent performance gap can be attributed to the broader dialectal coverage and the added functionality of voice cloning supported by our TTS model—capabilities that are absent in most commercial counterparts. Supporting a wide range of dialects and enabling personalized synthesis inevitably increases the modeling complexity, which in turn may lead to a modest degradation in performance on individual dialects.


\subsubsection{Ablation Study}
To demonstrate the superiority of our method, we conduct two ablation studies on four dialects.

The first evaluates the impact of the MoE. We compare a model with MoE against a baseline without it. The Second evaluates the impact of phonetic representation, we compare models using IPA and pinyin as input, both with the MoE architecture enabled. According to the results in Table \ref{tab:ablation}, it effectively proves that adding MoE and adopting IPA will achieve better results.

\section{Conclusion}
\label{sec:conclu}

We present a unified, open-source framework for low-resource dialect TTS, leveraging IPA-based modeling, dialect-aware MoE, and parameter-efficient adaptation to address data scarcity and phonetic variation. Datasets and code are released to ensure reproducibility. The framework provides a scalable foundation extendable to other language families. Future work will expand it with higher-quality data and broader dialect coverage, advancing open-data-driven multilingual speech synthesis.

\vfill\pagebreak


\bibliographystyle{IEEEbib}
\bibliography{strings,refs}

\end{document}